\documentclass[aps,manuscript,showpacs,showkeys]{revtex4}
\usepackage{graphicx}
\usepackage{epsfig}  
\usepackage{epsf}    
\usepackage{dcolumn}
\usepackage{bm}
\usepackage{dcolumn}
\def\be{\begin{equation}}
\def\ee{\end{equation}}
\def\bea{\begin{eqnarray}}
\def\eea{\end{eqnarray}}
\def\gsim{\ \rlap{\raise 2pt\hbox{$>$}}{\lower 2pt \hbox{$\sim$}}\ }
\def\lsim{\ \rlap{\raise 2pt\hbox{$<$}}{\lower 2pt \hbox{$\sim$}}\ }
\def\dslash{\kern-4pt \not{\hbox{\kern-2pt $\partial$}}}
\def\pslash{\not{\hbox{\kern-2pt p}}}

\def\nova{{NO$\nu$A}}

\def\pmue{{{\rm P_{\mu e}} }}

\def\dcp{\delta_{CP}}

\begin{document}


\title{Physics Potential of a 2540 Km Baseline Superbeam Experiment
 }


\author{Aniket Joglekar\footnote{Address after August 1, 2010: 
Department of Physics, University of Chicago, Chicago, IL, USA}}
\affiliation{
Department of Physics, Indian Institute of Technology Bombay,
Mumbai 400076, India}
 
\author{Suprabh Prakash}
\affiliation{
Department of Physics, Indian Institute of Technology Bombay,
Mumbai 400076, India}
 
\author{Sushant K. Raut}
\email[Email Address: ]{sushant@phy.iitb.ac.in}
\affiliation{
Department of Physics, Indian Institute of Technology Bombay,
Mumbai 400076, India}
 
\author{S. Uma Sankar}
\affiliation{
Department of Physics, Indian Institute of Technology Bombay,
Mumbai 400076, India}
\date{\today}
\begin{abstract}
We study the physics potential of a neutrino superbeam experiment 
with a 2540 km baseline. We assume a neutrino beam similar
to the NuMI beam in medium energy configuration. We consider a 
100 kton totally active scintillator detector at a 7 mr off-axis 
location. We find that such a configuration has outstanding 
hierarchy discriminating capability. In conjunction with the 
data from the present reactor neutrino experiments, it can 
determine the neutrino mass hierarchy at $3 \sigma$ level in less than
5 years, if $\sin^2 2 \theta_{13} \geq 0.01$, running in the 
neutrino mode alone. As a stand alone experiment, with a 5 year 
neutrino run and a 5 year anti-neutrino run, 
it can determine non-zero $\theta_{13}$
at $3 \sigma$ level if $\sin^2 2 \theta_{13} \geq 7 \times 10^{-3}$
and hierarchy at $3 \sigma$ level if $\sin^2 2 \theta_{13} \geq
8 \times 10^{-3}$.
This data can also distinguish $\dcp=\pi/2$ from the CP conserving values
of $0$ and $\pi$, for $\sin^2 2 \theta_{13} \geq 0.02$.
\end{abstract}
\maketitle

\vskip 1cm
\renewcommand{\thefootnote}{\arabic{footnote}}
\setcounter{footnote}{0}

\section{Introduction}

During the past decade, great progress has been made in
the study of neutrinos. In particular, the mass-squared
differences and two of the mixing angles have been 
measured with good precision. Despite this progress,
the following properties of neutrinos are still unknown 
\begin{itemize}
\item
The absolute mass scale of the neutrinos,
\item
Dirac vs Majorana nature of neutrinos,
\item
The value of the mixing angle $\theta_{13}$,
\item
Mass pattern or mass hierarchy of the neutrinos,
\item
CP violation in neutrino sector and the value of 
the CP-violating phase $\dcp$,
\item
Deviation of $\theta_{23}$ from $\pi/4$.
\end{itemize}

KATRIN tritium decay experiment \cite{katrin1,katrin2} and various 
neutrinoless double beta decay experiments \cite{ndbd} 
currently underway will address the first two problems.
The next four problems fall under the purview of neutrino
oscillations and hence are expected to be solved by the
current and future neutrino oscillation experiments. There
are three efforts underway to determine the mixing angle
$\theta_{13}$ using reactor neutrino data. They are 
(a) Double Chooz, expected to start taking data this year 
\cite{dchooz} with sensitivity $\sin^2 2 \theta_{13} \geq 0.04$, 
(b) RENO \cite{reno} with sensitivity $\sin^2 2 \theta_{13} \geq 0.02$,
and (c) Daya Bay \cite{dayabay} with 
sensitivity $\sin^2 2 \theta_{13} \geq 0.01$. 
In addition, there are two accelerator experiments, T2K
and \nova, which can determine non-zero $\theta_{13}$.
Their sensitivities are $\sin^2 2 \theta_{13} \geq 0.01$ 
at $90\%$ confidence level for T2K \cite{t2k} and 
$\sin^2 2 \theta_{13} \geq 0.03$ at $3 \sigma$ for \nova\ \cite{nova}.
Thus it is hoped
that the determination of $\theta_{13}$ can be achieved 
by these experiments within the next decade. 

However, it was shown recently \cite{huber2009} that the
currently running or planned experiments will be able
to determine the mass hierarchy or $\dcp$ only for very
favourable values of $\theta_{13}$ and $\dcp$. The effect
of mass hierarchy and $\dcp$ are closely interlinked 
\cite{degeneracy1,degeneracy2,degeneracy3,degeneracy4,degeneracy5}. 
Disentangling these two effects requires considerable
effort and possibly data from many experiments 
\cite{twobaseline1,twobaseline2}.  
Thus the question of designing a neutrino beam-detector
configuration, which will enable us to determine hierarchy 
and $\dcp$ unambiguously, acquires great importance. 

There have been numerous proposals made to achieve the
above purpose. These include megaton size detectors 
\cite{brookhome1,brookhome2,hyperk}, 
a pair of detectors with different baselines but with the
same value of $L/E$ \cite{twobaseline1,twobaseline2} and detectors at the 
`magic' baseline
of $7500$ km \cite{magic1,magic2}. Recently we made a proposal of a shorter
magical baseline \cite{bnlhs}, which has exceptional capability
to determine mass hierarchy, independently of $\dcp$.
In this report, we study the full physics
potential of our proposal.

\section{Calculation}

As mentioned before, the unknowns to be determined from the 
neutrino oscillation experiments are  
the mixing angle $\theta_{13}$, the neutrino mass hierarchy, 
the CP violating phase $\dcp$ and the octant of $\theta_{23}$.
The $\nu_\mu \rightarrow \nu_e$ oscillation probability $\pmue$ 
is sensitive to all these unknowns. 
In this paper, we will not consider 
the problem of octant ambiguity of $\theta_{23}$. We will assume
the true value of $\theta_{23}$ to be its best fit value $\pi/4$. 
The expression for $\pmue$, in three
flavour oscillations including matter effects, is given as 
\cite{numu2nue1,numu2nue2,numu2nue3}
\begin{eqnarray}
\pmue & = & 
C_0 \frac{\sin^2 ((1-\hat{A}) \Delta)}{(1-\hat{A})^2} \nonumber \\
& + & \alpha \ C_1  \frac{\sin((1-\hat{A}) \Delta)}{(1-\hat{A})}
\ \frac{\sin(\hat{A}\Delta)}{\hat{A}} \nonumber \\
& + & \alpha^2 C_2 \frac{\sin^2(\hat{A}\Delta)}{\hat{A}^2},
\label{pmue}
\end{eqnarray} 
where $\Delta = (1.27 \Delta_{31} L/E)$, $\hat{A} = A/\Delta_{31}$
and $\alpha = \Delta_{21}/\Delta_{31} \approx \pm 0.04$, as given
by the solar and atmospheric neutrino data \cite{globalfits}.  
Here $L$ is in km, $E$ is in GeV and $\Delta_{31}$ is given in 
units of eV$^2$. 
The coefficients, $C_i$ are given by 
\begin{eqnarray}
C_0 & = & \sin^2 \theta_{23} \sin^2 2 \theta_{13} ~,\\
C_1 & = & \cos \theta_{13} \sin 2 \theta_{12} \sin 2 \theta_{13} 
\sin 2 \theta_{23} \cos (\Delta+\delta_{CP}) ~,\\
C_2 & = & \sin^2 2 \theta_{12} \cos^2 \theta_{23} ~.
\end{eqnarray} 
The matter effects are given by the matter term \cite{wolfenstein}
$A \ ({\rm in~eV^2}) = 0.76 \times 10^{-4} \ \rho \ ({\rm gm/cc}) 
\ E \ ({\rm GeV})\ $eV$^2$. $\rho$ is the 
density of the matter through which the neutrino propagates. 
$\Delta_{31}$ and hence $\Delta$ are
positive for normal hierarchy (NH) and negative for 
inverted hierarchy (IH), for both neutrinos and anti-neutrinos. 
$A$, on the other hand, is positive for neutrinos and negative
for anti-neutrinos. 
Therefore, $\hat{A}$ is positive for NH and negative for IH 
in the case of neutrinos, and vice versa for anti-neutrinos. 

 
The dependence of $\pmue$ on $\dcp$ and $\theta_{13}$ 
is seen explicitly in Eq.~(\ref{pmue}). In addition, 
the quantities $\hat{A}$ and $\Delta$ are sensitive to the 
neutrino mass hierarchy. Thus, this 
channel  which is sensitive to all the three unknowns mentioned above, 
is a good candidate to probe these unknown parameters. 
From Eq.~(\ref{pmue}) we see that $\pmue$ is large for  
$\Delta = (1.27 \Delta_{31} L/E) \approx \pi/2$. On the other hand, 
the matter term $\hat{A}$ becomes significant at large energies. 
Therefore, the matter term starts to play an important role in shaping 
the oscillation probabilities only for long baselines. 
For a baseline of $732$ km, and for fixed values of $\theta_{13}$ and
$\dcp$, it increases $\pmue$ by about $25 \%$ for NH and decreases it by the 
same amount for IH \cite{mnsu}. For longer baselines, the change
in the probability is even larger. 
At shorter baselines, the sensitivity of $\pmue$ to 
the matter term and hence the mass hierarchy, is low. 

\begin{figure}
\begin{center}
\epsfig{file=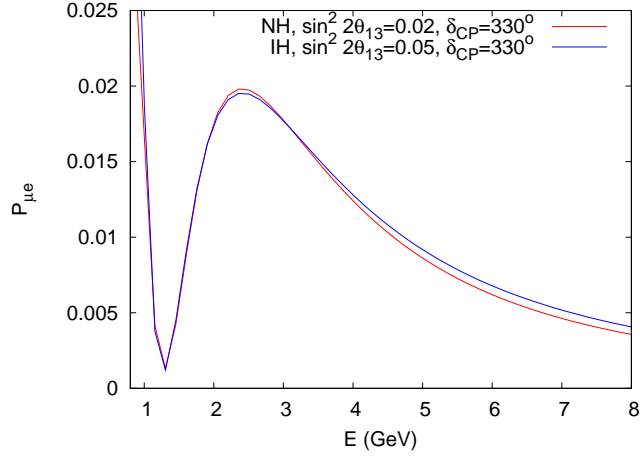,width=3.5in}
\caption{\footnotesize
Hierarchy-$\theta_{13}$ degeneracy in $\pmue$ at $1300$km. We see 
that $\pmue (NH, \sin^2 2 \theta_{13}=0.02; E) = 
\pmue (IH, \sin^2 2 \theta_{13}=0.05; E)$. The value of $\dcp$ is taken 
to be $330^\circ$ here.}
\label{fig1}
\end{center}
 \end{figure}

Since $\pmue$ depends on three unknowns, $\theta_{13}$,
mass hierarchy and $\dcp$, it is impossible to determine
any one of them by making a single measurement. Even if
one measures $\pmue$ as a function of energy, it is 
possible to get degenerate solutions 
\cite{degeneracy1,degeneracy2,degeneracy3,degeneracy4,degeneracy5}.
For example, it is
possible to have $\pmue (\theta_{13}, NH; E) = 
\pmue (\theta_{13}',IH; E)$, as illustrated in Fig~\ref{fig1}.
Such a degeneracy can be broken, if the value of $\theta_{13}$
can be determined from different experiments. Reactor
anti-neutrino experiments with $L \approx 1$ km, can determine
$\theta_{13}$ unambiguously. The currently planned experiments
can measure non-zero values as small as $\sin^2 2 \theta_{13}
\geq 0.01$ \cite{dchooz,reno,dayabay}. Thus we expect that this 
degeneracy can be lifted
by the results of reactor neutrino experiments.

\begin{figure}
\begin{center}
\epsfig{file=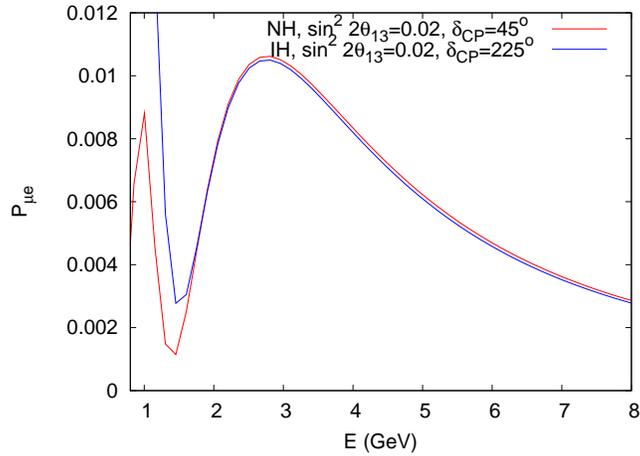,width=3.5in}
\caption{\footnotesize
Hierarchy-$\dcp$ degeneracy in $\pmue$ at $1300$km. We see 
that $\pmue (NH, \dcp=45^\circ; E) = 
\pmue (IH, \dcp=225^\circ; E)$ for $E > 1.5$ GeV. 
The value of $\sin^2 2\theta_{13}$ is 
taken to be $0.02$ here.}
\label{fig2}
\end{center}
 \end{figure}

Even if the ambiguity due to the value of $\theta_{13}$  
is lifted, there is a further ambiguity between hierarchy
and $\dcp$. This ambiguity arises due to the following
reason. If we replace $\nu \rightarrow \bar{\nu}$, we have $\dcp
\to -\dcp$ and $\hat{A} \to - \hat{A}$. However, if we 
flip the hierarchy, we have $\hat{A} \to - \hat{A}$ and 
$\Delta \to - \Delta$. Hence, the change in $\pmue$ due to
hierarchy flip can be compensated by changing $\dcp$. 
This ambiguity is particularly important for $\sin^2 2
\theta_{13} \leq 0.02$. Therefore,
even if we know the value of $\theta_{13}$, it is possible
to get degenerate solutions of the kind $\pmue (NH, \dcp; E)
= \pmue (IH, \dcp'; E)$ over a wide range of energies,
as illustrated in Fig.~\ref{fig2}. 
We are led to consider various improvisations to overcome 
this problem. 

\begin{figure}
\begin{center}
\epsfig{file=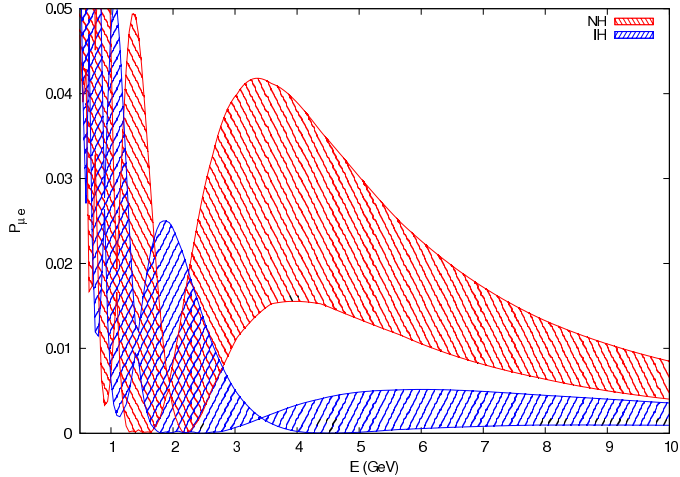,width=3.5in}
\caption{\footnotesize
$\pmue$ as a function of $E$ for $L=2540$ Km and $\sin^2 2 \theta_{13} = 0.02$. 
$\pmue$ is plotted for both NH and IH, each for the full range of values 
of $\dcp$.}
\label{fig3}
\end{center}
 \end{figure}

In Eq.~(\ref{pmue}), only the $C_1$ term contains $\dcp$.
The hierarchy-$\dcp$ ambiguity can be lifted if the term
containing $C_1$ can be chosen to be zero. One simple way
to do it is to choose $\hat{A} \Delta = \pi$ \cite{smirnov}. This 
condition can be achieved for both NH and IH simultaneously
and gives us an energy independent condition $L \approx 7500$ km.
This is the famous magic baseline condition and it has been
extensively studied \cite{magic1,magic2}. In \cite{bnlhs} it was noted that the
$C_1$ term can be made zero for IH only by choosing $(1-\hat{A})
\Delta = - \pi$. This makes $\pmue(IH)$ independent of
$\dcp$ and also very small $(\leq \alpha^2 \approx 0.002)$. If we choose
$(1-\hat{A})\Delta = \pi/2$ for NH, then $\pmue(NH)$ will be
quite siginificant for all values of $\dcp$. Solving the above pair
of simultaneous equations, we obtain $L=2540$ km and $E = 3.3$ GeV.
For this baseline, at this energy, there is a clear separation 
between $\pmue(NH)$ and $\pmue(IH)$ 
as illustrated in Fig.~\ref{fig3}. This is true for values of $\theta_{13}$
at least as small as those detectable by the current reactor experiments. 
In addition, $\pmue(NH)$ 
retains its dependence on $\dcp$ unlike in the 
magic baseline case. This, in turn, can be helpful in measuring
$\dcp$ \cite{bimagic}.

In our calculations, we have considered a superbeam experiment with 
a baseline of $2540$ km. Note that the distance from Brookhaven Laboratory 
to Homestake mine \cite{brookhome1,brookhome2} and that from CERN to Pyhasalmi 
\cite{peltoniemi09} are close to our shorter `magical' baseline. Since our 
energy range of interest is around $3-4$ GeV, the neutrino source must 
have unoscillated event spectrum peaking in this energy range. 
The NuMI beam in the medium energy option has an unoscillated event 
spectrum that peaks at $3.5$ GeV for locations $7$ mr off the beam axis 
\cite{nova}. 

\section{Results}

We have done all our calculations using the software package 
GLoBES \cite{globes1,globes2}. The source is taken to be a NuMI like
beam in the medium energy configuration. The detector is 
assumed to be a 100 kton totally active scintillator detector
whose capabilities are similar to \nova. Such a detector is 
placed at a distance 2540 km from the source at a 7 mr off-axis
location. The fluxes for this location were calculated using 
the program from \cite{messier}. The beam power is taken to be
such that it corresponds to $10\times10^{20}$ POT/yr. This is
about $30\%$ more than the design power for \nova.
Thus we have an exposure of $1000 \times 10^{20}$ kton-POT/yr.
This is about $10$ times larger than the exposure for \nova.
The baseline in our proposal is thrice as long as that of 
\nova. Hence the factor $10$ increase in exposure makes the 
data to have the same statistical weight as that of \nova.

We have computed the $\nu_e$ appearance spectrum in the detector
for the energy range from 250 MeV to 10 GeV, in bins of 250 MeV 
width. A Gaussian energy smearing function with a width of 
$0.1\sqrt{E \ (GeV)}$ has been assumed for these events. We have 
considered four different types of events in our analysis: signal 
($\nu_\mu \rightarrow \nu_e$) events, beam background events
(that is, events caused by intrinsic $\nu_e$ component of the
beam), background events due to misidentified muons and the
background events due to neutral current interactions. Energy independent 
cuts identical to those for \nova\ have been imposed on these channels, 
which substantially enhance the signal to background ratio
\cite{nova,globesnova}. 

In our calculations, the solar parameters $\sin^2 \theta_{12}=0.304$ 
and $\Delta_{21} =7.65 \times10^{-5}eV^2$ have been kept fixed throughout
\cite{globalfits,kamland1,kamland2}. 
The values of $\Delta_{31}$ and $\theta_{23}$ are taken from 
three flavour oscillation fit of the atmospheric \cite{skatm} 
and MINOS data 
\cite{minos}. We take the true value of $\sin^2 2 \theta_{23}$ 
to be its best fit value $1$.
We assumed an error of $2\%$ on this measurement.  
Hence the test values of 
$\theta_{23}$ are to be chosen to satisfy the constraint
$0.94 \leq \sin^2 2 \theta_{23} \leq 1$. 
Thus we have the range 
$38^\circ \leq \theta_{23} \leq 52^\circ$. The situation 
regarding $\Delta_{31}$ is a bit more complicated. The most 
accurate measurement of the larger mass-squared difference is 
given by the MINOS experiment \cite{minos}. 
If the results of MINOS
are interpreted in terms of three flavour oscillations, 
the effective mass-squared difference measured by it  
is not $|\Delta_{31}|$ but is the magnitude of  
a linear combination of $\Delta_{31}$ and $\Delta_{21}$ 
\cite{degouvea,parke2005}. 
It is shown \cite{parke2005} that the linear combination is of the form
$\Delta_{31} + \Delta_{21} f(\theta_{ij},\dcp)$, where 
\begin{equation}
f(\theta_{ij},\dcp) = \cos^2\theta_{12} - \cos\dcp\ \sin\theta_{13}\ 
\sin2\theta_{12}\ \tan\theta_{23}.
\end{equation}
In computing $\pmue$(NH) and
$\pmue$(IH) we must relate the corresponding values of $\Delta_{31}$
via 
\begin{equation} 
\Delta_{31} \ {\rm (IH)} = -  \Delta_{31} \ {\rm (NH)} - 
2 \Delta_{21} f (\theta_{ij},\dcp),
\label{d31relation}
\end{equation}
rather than as $\Delta_{31} \ {\rm (IH)} = - \Delta_{31} \ {\rm (NH)}$.
If we were to take 
$\Delta_{31} \ {\rm (IH)} =-\Delta_{31} \ {\rm (NH)}$, 
then we
would get a significant {\it fake hierarchy sensitivity} in the 
$\nu_\mu$ disappearance channel, even for
very small values of $\theta_{13}$. 
In our calculations, we have related  
$\Delta_{31} \ {\rm (NH)}$ and $\Delta_{31} \ {\rm (IH)}$
via Eq.(\ref{d31relation}), and used the best fit value 
$\Delta_{31}\ {\rm (NH)}=2.4 \times10^{-3}eV^2$, with an error of 5\%. 
The systematic errors include a 5\% normalization 
error and 2.5\% tilt error on both the signal and the background. We have 
also taken into consideration a 5\% error in the density profile of the 
earth. 

First we computed the hierarchy sensitivity of this setup,
running it in the neutrino mode only. In this part of the calculation,
we limited ourselves to the range $0.01 \leq \sin^2 2 \theta_{13} \leq 0.1$.
That is, we assumed that this parameter will be determined
by one of the current experiments. First we assumed NH to be
the true hierarchy and computed the $\chi^2$ for distinguishing
this from IH, as a function of $\sin^2 2 \theta_{13}$. In 
computing the $\chi^2$, we combined the results of our proposed
set up with those of 
the reactor experiments and of T2K and \nova\ 
\cite{t2k,nova,globesnova,globesreactor,globest2k1,globest2k2}. 
The data from the
reactor experiments provide a `prior' on the 
values of $\sin^2 2 \theta_{13}$. In computing $\chi^2$, we
have marginalized over the range $0.01 \leq \sin^2 2 \theta_{13} \leq 0.1$.
We have varied the CP phase over the full range $-\pi \leq \dcp \leq \pi$
in computing the event rates both for NH and for IH. 
We have then repeated the above calculation, this time assuming
IH to be the true hierarchy.  
The results of these 
calculations are tabulated in Table \ref{exposure}. We indicate here the 
exposure time (in years) required for hierarchy discrimination up to a 
$3\sigma$ confidence level as a function of $\sin^2 2 \theta_{13}$. 
\newline

\begin{table}[tbh]
\begin{tabular}{|c|c|c|}
\hline
$\sin^2 2\theta_{13}$ (true) & Exposure time(NH) & Exposure time(IH) \\ \hline
0.10 & 0.022  & 0.048 \\
 \hline
0.09 & 0.026 & 0.057 \\
 \hline
0.08 & 0.031 & 0.068 \\
 \hline
0.07 & 0.040 & 0.082 \\
 \hline
0.06 & 0.051 & 0.105 \\
 \hline
0.05 & 0.070 & 0.137 \\
 \hline
0.04 & 0.104 & 0.195 \\
 \hline
0.03 & 0.180 & 0.420 \\
 \hline
0.02 & 0.425 & 2.600 \\
 \hline
0.01 & 2.950 & 4.800 \\
 \hline
\end{tabular}
\caption{Exposure time in years required for $3\sigma$ hierarchy 
discrimination, for $1000\times10^{20}$kton-POT/yr. The 
second (third) column shows the results if NH (IH) is the true hierarchy.
It is assumed that the experiment runs in neutrino mode only.}
\label{exposure}
\end{table}
From the results of the table, it is worth remarking that,
the present 15 kton \nova\ detector, with its current design
luminosity, can determine the neutrino mass hierarchy, independently
of $\dcp$, in a five year neutrino run, 
if $\sin^2 2 \theta_{13} \geq 0.03$, were the
detector placed at a distance of 2540 km with an appropriately
designed beamline.

\begin{center}
\begin{figure}[h]
\begin{tabular*}{1.0\textwidth}{@{}c@{}c}
\epsfig{file=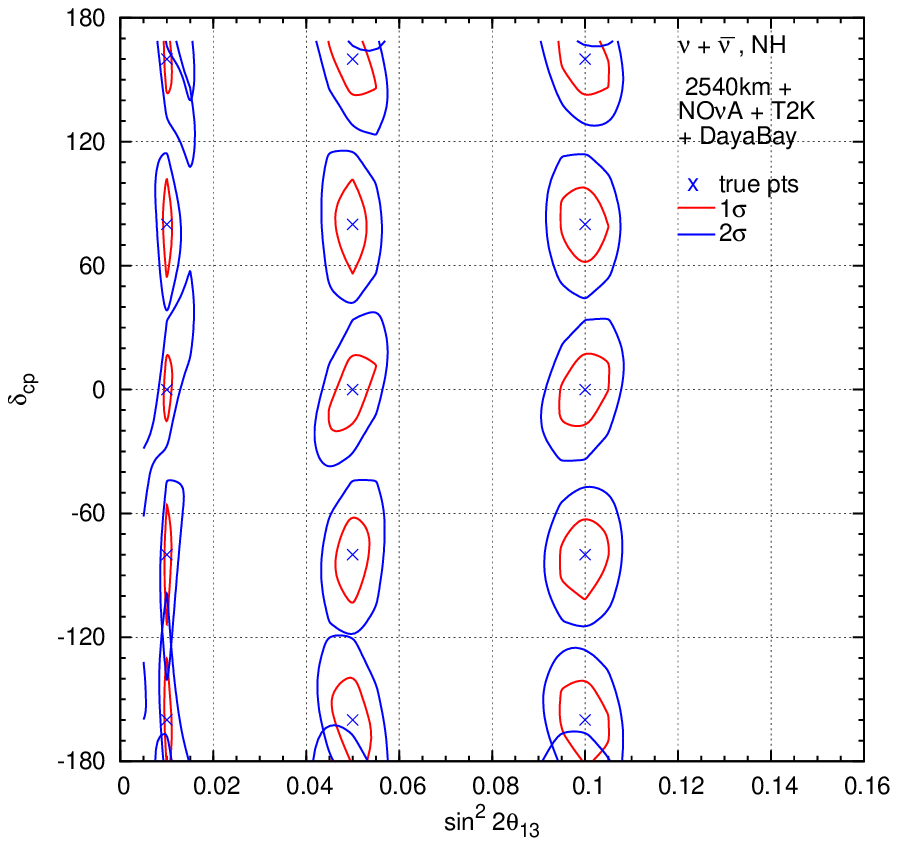,width=3.5in} & 
\epsfig{file=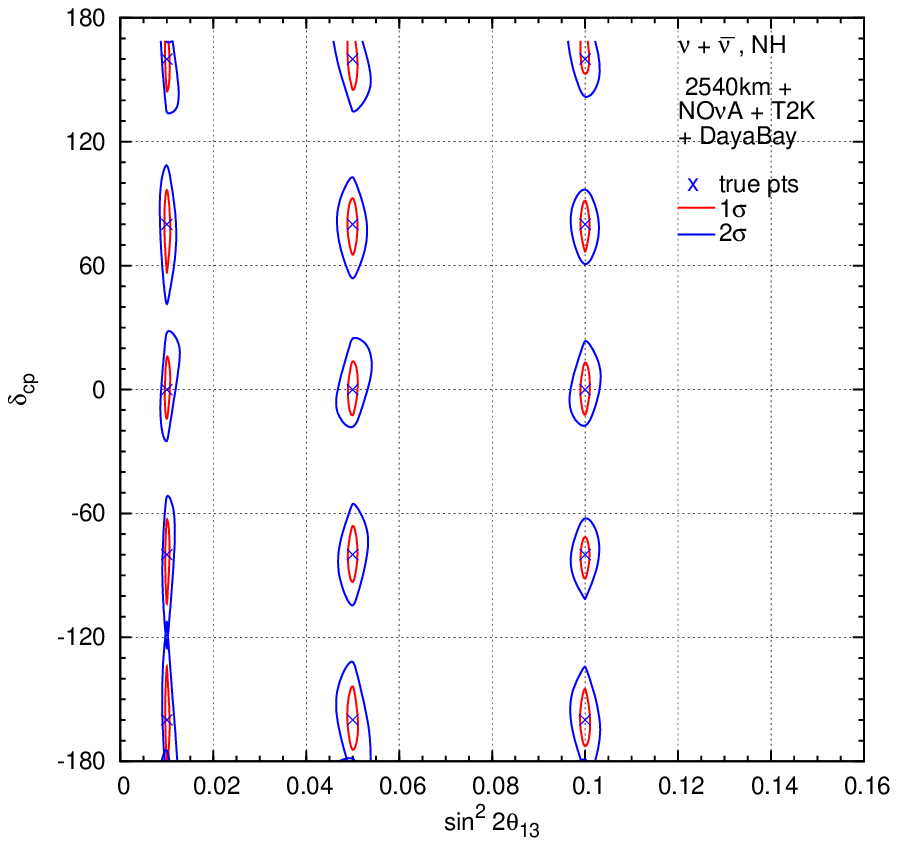,width=3.5in}
\end{tabular*}
\caption{\footnotesize
$1\sigma$ and $2\sigma$ contours for sensitivity in the $\sin^2 2 
\theta_{13}-\dcp$ 
plane. The plot on the right is without systematics.}
\label{peanuts}
 \end{figure}
\end{center}

Next, we have calculated the ability of our setup (in conjunction with 
T2K, \nova\ and the reactor experiments) to determine $\dcp$ as a 
function of $\sin^2 2 \theta_{13}$. In addition to the data from a 
five year neutrino run, we assume data from a five year anti-neutrino
run. We have chosen fifteen different
pairs of true values of $\sin^2 2 \theta_{13}$ and $\dcp$ given by
$\sin^2 2 \theta_{13} =0.01,0.05,0.10$ and $\dcp=-160^\circ,-80^\circ,0^\circ,
80^\circ,160^\circ$. We assumed normal hierarchy and computed the  
allowed $1\sigma$ and $2\sigma$ regions in the $\sin^2 2\theta_{13}-\dcp$ 
plane. 
In Figure \ref{peanuts}, we have plotted the outer contours of these
allowed regions, corresponding to the fifteen different true parameter sets. 
The panel on the left gives the results including the systematic
errors, whereas the panel on the right shows the allowed contours
without the systematic errors. From the right panel, we see that
the allowed regions are very close to the true values and the value
of $\dcp$ can be determined with an uncertainty of about $10^\circ$.
The value of $\sin^2 2 \theta_{13}$ is rather tightly constrained
by the data from reactor neutrino experiments. However, the inclusion
of systematic errors makes the allowed regions considerably larger.
$\sin^2 2 \theta_{13}$ is allowed to deviate from its true value
by about $0.005$ and the uncertainty in determining $\dcp$ rises
to about $20^\circ$.

We have also computed the physics capabilities of the proposed setup 
as a stand alone experiment. As mentioned before, in the previous
calculations, we have taken a lower limit of $0.01 \leq \sin^2 2 \theta_{13}$.
The reason for this is, that for values of $\theta_{13}$ above this 
limit, the data from the reactor experiments provide a `prior'.
In the calculations discussed below, we did not impose this 
lower limit on $\theta_{13}$ but allowed it to be as small as
$\sin^2 2 \theta_{13} \simeq 10^{-4}$. 

In computing the results discussed below, we have assumed a 
5 year neutrino and a 5 year anti-neutrino run. Based on
this data, we computed the sensitivity of this set up to 
\begin{itemize}
\item
measure non-zero $\theta_{13}$
\item
determine hierarchy
\item 
rule out the CP conserving value of $\dcp = 0 \ {\rm or} \ \pi$.
\end{itemize}
 
\begin{center}
\begin{figure}[h]
\epsfig{file=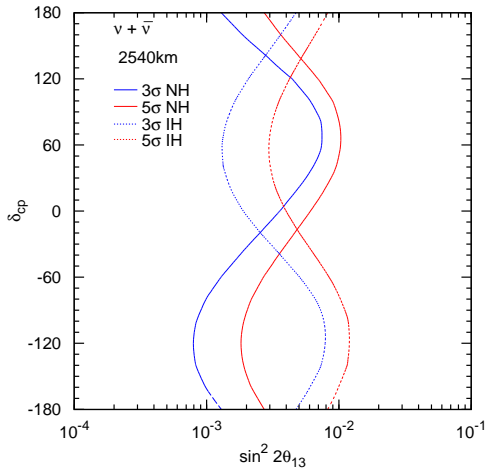,width=4.5in}  
\caption{\footnotesize
$3\sigma$ and $5\sigma$ contours in the $\sin^2 2 \theta_{13}-\dcp$ 
plane for excluding $\theta_{13}=0$. 
}
\label{exclth13}
 \end{figure}
\end{center}

In figure \ref{exclth13}, we plot the true values of $\sin^2 2 
\theta_{13}$ for which this parameter can be distinguished from
$0$. We do this as a function of the true value of $\dcp$ for 
both NH and IH. 
If NH is the true hierarchy, then values of $\sin^2 2 \theta_{13}$
as small as $8 \times 10^{-4}$ can be measured at $3 \sigma$ level,
for advantageous values of $\dcp$ (between $-\pi/3$ to $-2\pi/3$).
For adverse values of $\dcp$ (between $\pi/6$ to $\pi/2$), however,
$\theta_{13}$ can be distinguished from $0$ only for $\sin^2 2 
\theta_{13} \geq 7 \times 10^{-3}$. If IH is the true hierarchy, the results
are similar, except for the fact that the advantageous and adverse
values of $\dcp$ are exchanged. 

\begin{center}
\begin{figure}[h]
\epsfig{file=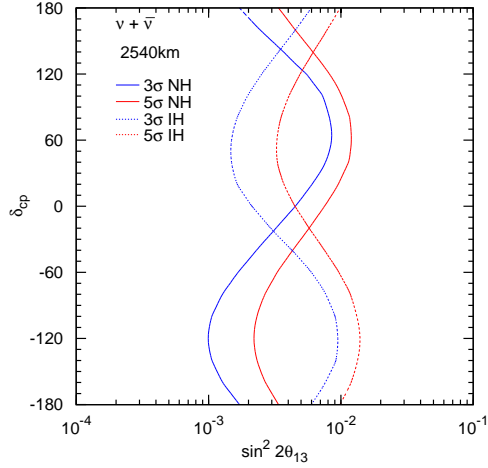,width=4.5in}  
\caption{\footnotesize
$3\sigma$ and $5\sigma$ contours in the $\sin^2 2 \theta_{13}-\dcp$ 
plane for excluding the `wrong' hierarchy. 
}
\label{exclhier}
 \end{figure}
\end{center}

In figure \ref{exclhier}, we plot the sensitivity of this setup for 
distinguishing the mass hierarchy as a function of the true value of
$\sin^2 2 \theta_{13}$ and $\dcp$. Once again, for advantageous values
of $\dcp$, the mass hierarchy can be determined at $3 \sigma$ 
for $\sin^2 2 \theta_{13} \geq 8 \times 10^{-4}$, whereas for 
adverse values, it can be determined only for values which are about 
an order of magnitude larger.

Here we must insert a note of caution. Figures~\ref{exclth13} 
and~\ref{exclhier} give the sensitivity of our proposed setup
as a function of the true value of $\dcp$. However, the true
value of $\dcp$ is unlikely to be known from any of the current
experiments. Hence, while making a statement regarding the 
sensitivity of our setup, we must consider the most advserse
value of $\dcp$. Thus, a five year neutrino and a five year
anti-neutrino run in our setup can measure, {\it independently of $\dcp$}, 
$\sin^2 2 \theta_{13}$ if it is larger than $7 \times 10^{-3}$ 
and it can determine hierarchy
for a similar lower limit on $\theta_{13}$.

\begin{center}
\begin{figure}[h]
\epsfig{file=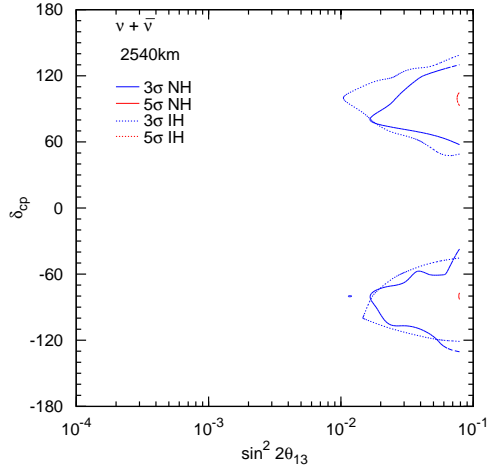,width=4.5in}  
\caption{\footnotesize
$3\sigma$ and $5\sigma$ contours in the $\sin^2 2 \theta_{13}-\dcp$ 
plane for excluding 
CP conservation. 
}
\label{exclcpnv}
 \end{figure}
\end{center}

Finally, figure \ref{exclcpnv} shows the ability of the setup to 
exclude the CP-conserving case, $\dcp=0,\pi$. We find that the 
maximal phase $\dcp = \pm \pi/2$ can be distinguished from 
$\dcp = 0, \pi$ for $\sin^2 2 \theta_{13} \geq 0.01$. But for 
other values, discrimination from CP conserving case is possible
only for larger values of $\theta_{13}$. In fact, for $\dcp$
outside the two ranges $-2\pi/3 \ {\rm to} \ -\pi/3$ and 
$\pi/3 \ {\rm to} \ 2 \pi/3$, the discrimination from CP conserving
case is not possible even for the largest allowed value
of $\theta_{13}$. 

H2B collaboration \cite{h2b1} have considered a neutrino superbeam 
experiment with a baseline in excess of 2000 km. Their setup
envisaged directing a 4 MW neutrino superbeam, with peak flux
at $E_\nu = 4$ GeV, from Tokai in Japan to a 100 kton water
Cerenkov detector at Beijing, 2100 km away. Such a set up
is close to satisfiying the shorter magic baseline conditions derived in
section-II. They found that
\cite{h2b2,h2b3} such a setup, running only in neutrino
mode for five years, can determine mass hierarchy for all values of $\dcp$,
if $\sin^2 2 \theta_{13} \geq 0.02$.

There was a previous study of a superbeam
proposal with 2540 km baseline \cite{brookhome1,brookhome2}. In this
proposal, a wide band neutrino beam, with $0.5 \leq E_\nu
\leq 7$ GeV energy range from Brookhaven National Laboratory
is directed at a 500 kton water Cerenkov detector at 
Homestake mine. The beam is expected to be produced by
28 GeV protons with 1 MW intensity. This corresponds
to $22 \times 10^{20}$ POT/yr assuming one year contains 
$10^7$ sec running time \cite{dusel1,dusel2}.
The detector is assumed to be along the
beamline, rather than at an off-axis location, which leads
to the rather wide range in the energy. With a 5 year run
in neutrino mode only, this setup can set a $90\%$ confidence 
level upper limit $\sin^2 2 \theta_{13} \leq 0.005$ if 
NH is the true hierarchy and $\sin^2 2 \theta_{13} \leq 0.02$
if IH is the true hierarchy. It can also determine $\dcp$ 
with about a $25\%$ uncertainty.

A few years ago, a US long baseline neutrino study group 
advocated a 1300 km baseline experiment to determine the
unknown neutrino parameters from oscillations \cite{dusel1,dusel2}. 
They considered
a wide band neutrino beam with $0.25 \leq E_\nu \leq 7$ GeV 
energy range directed from Fermilab to a 300 kton water Cerenkov 
detector at Homestake mine. They assumed a neutrino run with
$30 \times 10^{20}$ POT with a similar anti-neutrino run. 
The exposure of this setup, in POT-kton, is twice that
of the setup we considered. The CP-violation discovery potential 
of this setup is better but the hierarchy determination potential 
of our proposal is better \cite{huberkopp}.

Recently, there was another study which considered
additional magical properties of the 2540 km baseline \cite{bimagic}.
In that study, the following set up was considered. The neutrino
source is a low energy neutrino factory with a flux of $5 \times
10^{21}$ muons/year with the muon energy of 5 GeV. The detector
is a 25 kton totally active scintillator detector and it is 
assumed that this setup will run with positive muons for 
2.5 years. With these beam and detector specifications, 
the authors showed that $\theta_{13}$ can be measured for
values $\sin^2 2 \theta_{13} \geq 4 \times 10^{-3}$ and the
hierarchy can be determined for $\sin^2 2 \theta_{13} \geq
8 \times 10^{-3}$. This set up seems to have a good ability 
to discover non-zero $\dcp$.


\section{Conclusion}
In an earlier paper, we demonstrated that a neutrino superbeam 
experiment with a 2540 km baseline is particularly well suited
to determine neutrino mass hierarchy, independently of the value
of $\dcp$. In this paper, we performed a full-fledged study of
the physics  potential of such an experiment. We considered a
100 kton totally active scintillator (\nova\ like) detector with 
the neutrino source being a NuMI-like beam in medium energy option with 
$10 \times 10^{20}$ POT/yr. We performed the analysis using
GLoBES software, where all the appropriate background events
have been included and background suppression factors relevant
for \nova\ are imposed. We demonstrate that a five year neutrino run in 
this setup, together with data from reactor neutrino experiments, 
can determine the neutrino hierarchy {\it independently of $\dcp$}
for $\sin^2 2 \theta_{13} \geq 0.01$. This statement holds 
irrespective of whether the true hierarchy is NH or IH. With
additional data from a five year anti-neutrino run, this setup
is capable of determining $\dcp$ with an uncertainty of about
$20^\circ$.

If this setup is considered by itself, then the data from a
five year neutrino run plus a five year anti-neutrino run
can measure non-zero $\theta_{13}$ and determine neutrino
mass hierarchy for $\sin^2 2 \theta_{13} \geq 7 \times 10^{-3}$.
These statements hold true for all possible values of $\dcp$.
It is also possible to obtain signals for CP violation, for
moderately large values of $\sin^2 2 \theta_{13}$.

\noindent
{\bf Acknowledgement} We thank Ravi Shanker Singh for collaboration 
on an earlier paper where the magical properties of 2540 km baseline
were first discussed. We thank 
Srubabati Goswami for numerous discussions regarding various aspects
of 2540 km baseline and Patrick Huber for discussions on running of
GLoBES.  

\end{document}